\documentclass[aps,prl,reprint,superscriptaddress,longbibliography]{revtex4-1}
\usepackage{float}
\usepackage{graphicx}
\usepackage{epsfig}
\usepackage{color}
\usepackage{soul}
\usepackage{amsmath,amssymb}

\begin{document}

\title{Active Matter Alters the Growth Dynamics of Coffee Rings}

\author{Tugba Andac}
\affiliation{Soft Matter Lab, Department of Physics, Bilkent University, Ankara 06800, Turkey}

\author{Pascal Weigmann}
\affiliation{Soft Matter Lab, Department of Physics, Bilkent University, Ankara 06800, Turkey}

\author{Sabareesh K. P. Velu}
\affiliation{Soft Matter Lab, Department of Physics, Bilkent University, Ankara 06800, Turkey}
\affiliation{University of Information Science and Technology ``St. Paul The Apostle", Ohrid 6000, Macedonia}

\author{Er\c{c}a\u{g} Pin\c{c}e}
\affiliation{Soft Matter Lab, Department of Physics, Bilkent University, Ankara 06800, Turkey}

\author{Agnese Callegari}
\affiliation{Soft Matter Lab, Department of Physics, Bilkent University, Ankara 06800, Turkey}

\author{Giorgio Volpe}
\affiliation{Department of Chemistry, University College London, 20 Gordon Street, London WC1H 0AJ, United Kingdom}

\author{Giovanni Volpe}
\email{giovanni.volpe@physics.gu.se}
\affiliation{Department of Physics, University of Gothenburg, SE-41296 Gothenburg, Sweden}
\affiliation{Soft Matter Lab, Department of Physics, Bilkent University, Ankara 06800, Turkey}

\date{\today}

\begin{abstract}
How particles are deposited at the edge of evaporating droplets, i.e. the {\em coffee ring} effect, plays a crucial role in phenomena as diverse as thin-film deposition, self-assembly, and biofilm formation. Recently, microorganisms have been shown to passively exploit and alter these deposition dynamics to increase their survival chances under harshening conditions. Here, we show that, as the droplet evaporation rate slows down, bacterial mobility starts playing a major role in determining the growth dynamics of the edge of drying droplets. Such motility-induced dynamics can influence several biophysical phenomena, from the formation of biofilms to the spreading of pathogens in humid environments and on surfaces subject to periodic drying. Analogous dynamics in other active matter systems can be exploited for technological applications in printing, coating, and self-assembly, where the standard coffee-ring effect is often a nuisance.
\end{abstract}

\maketitle

\begin{figure}[h!]
\includegraphics[width=0.5\textwidth]{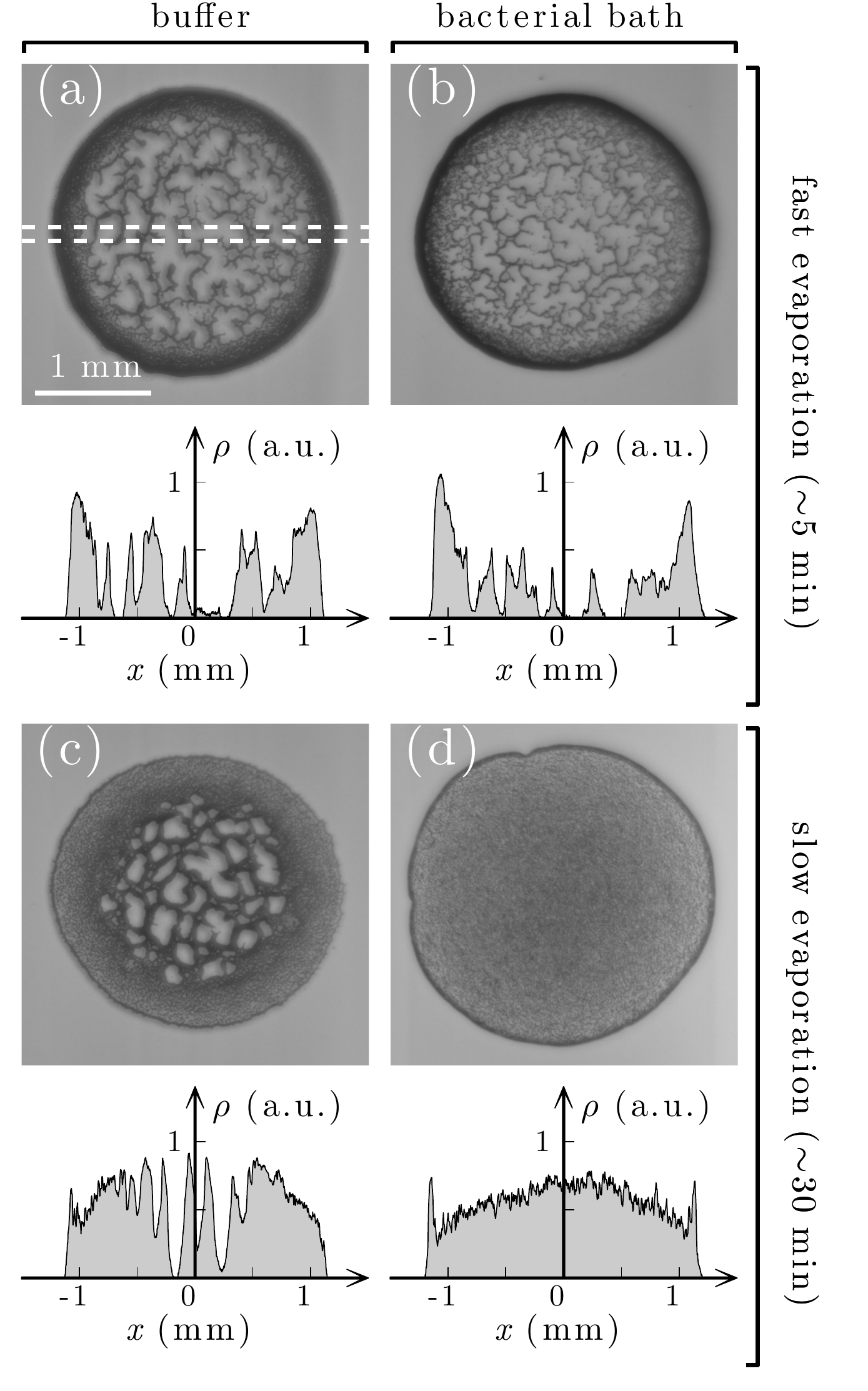}
\caption{
\textbf{Stain at the end of the evaporation with and without bacteria.}
Stain left behind by a droplet (a,b) after fast evaporation ($\sim$5 minutes) and (c,d) after slow evaporation ($\sim$30 minutes). The droplet is made of a buffer solution containing colloidal particles (polystyrene microsphere, diameter $2R = 3.00 \pm 0.07\,{\rm \mu m}$) (a,c) without and (b,d) with motile bacteria (\emph{E. coli}). The plot below each panel shows the optical density of the deposit along one droplet's diameter (dashed lines in (a)) as calculated from the image inverted gray scale. (a,b) For fast evaporation, both stains share similar features. (c,d) For slow evaporation, the stain of the droplet containing bacteria features higher uniformity than all other cases. See also supplementary video 1.}
\label{fig1}
\end{figure}

\begin{figure*}
\includegraphics[width=\textwidth]{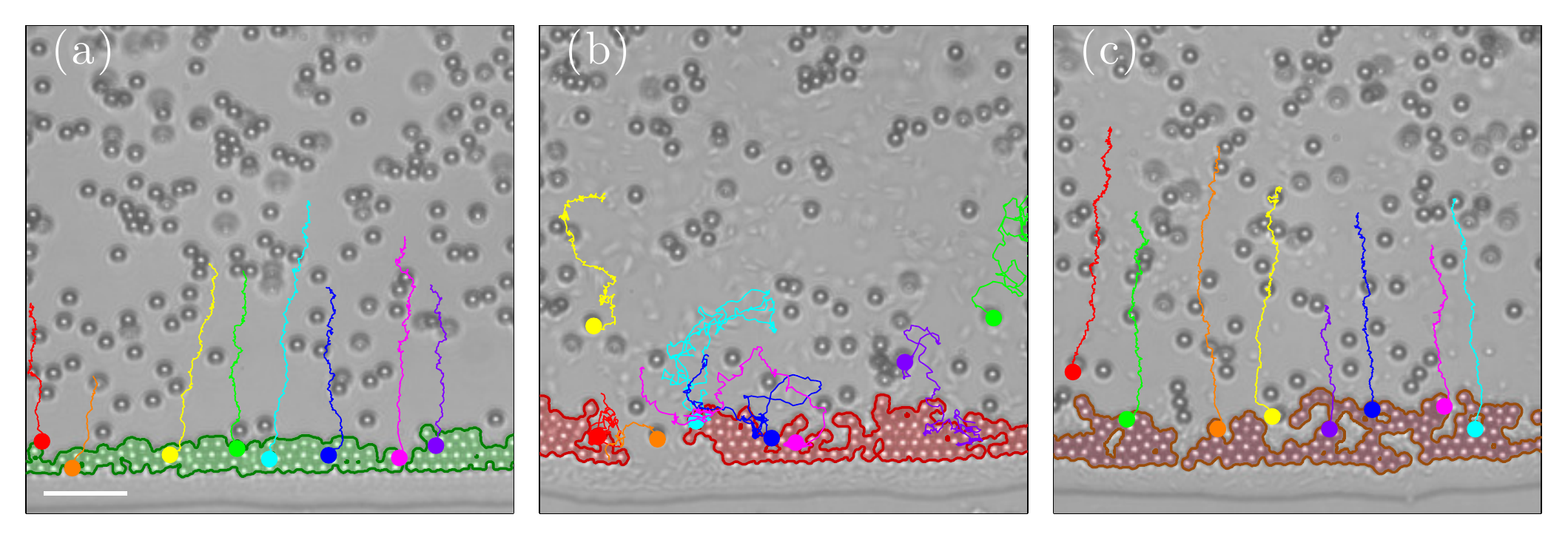}
\caption{
\textbf{Growth dynamics at the edge of drying droplets.}
Edge of a drying droplet of motility buffer containing (a) only colloidal particles, (b) colloidal particles and motile bacteria, and (c) colloidal particles and non-motile bacteria.
The shaded area represents the already formed border, and the solid lines show the trajectories of some particles recorded over the preceding $60\,{\rm s}$.
In (b), because of the presence of motile bacteria, the particle trajectories are more complex and also feature events where a particle escapes the boundary after having reached it.
The scalebar corresponds to $20\,{\rm \mu m}$. See also supplementary video 2.
}
\label{fig2}
\end{figure*}

\begin{figure*}
\includegraphics[width=\textwidth]{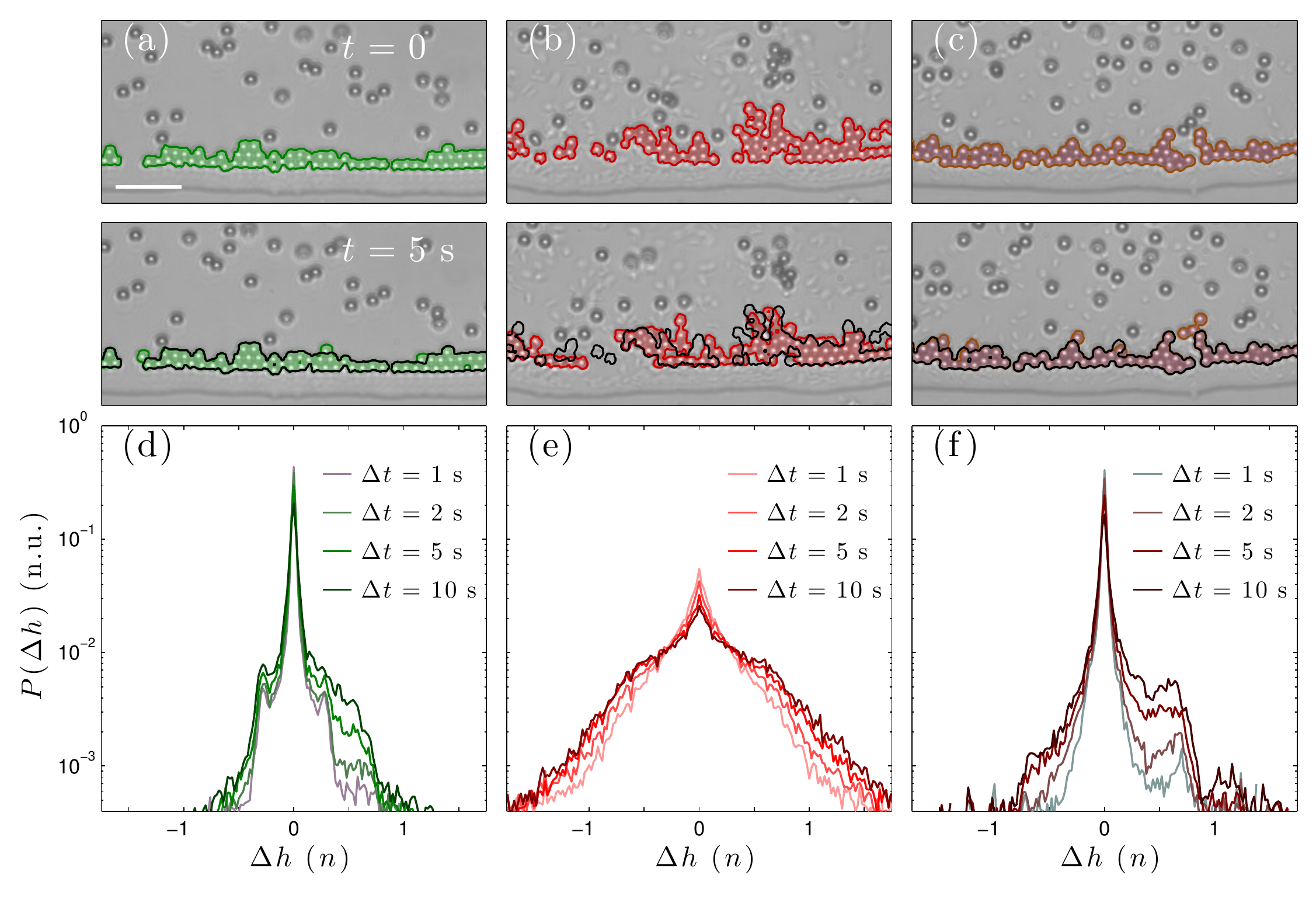}
\caption{
\textbf{Edge growth.}
(a-c) Two snapshots of the droplet's edge separated by a time lag $\Delta t = 5\,{\rm s}$ for the cases of a drying droplet containing (a) only colloidal particles, (b) colloidal particles and motile bacteria, and (c) colloidal particles and non-motile bacteria. The shaded areas represent the border already formed at each time. The black solid lines in the bottom panels reproduce the border in the respective top panels to ease comparison. (a) Without bacteria and (c) with non-motile bacteria, the boundary grows only because new particles deposit and no particles detach from it over time, while, (b) with motile bacteria, the boundary changes shape in a more dynamic way as particles, after arriving at the boundary, keep moving and can even detach from it. The scalebar corresponds to $20\,{\rm \mu m}$. (d-f) Distributions of the change of the boundary height (measured in numbers of particle layers in a close-packed border, so that $\Delta h = 1$ means that the border has grown by one particle) for time intervals $\Delta t = 1$, $2$, $5$ and $10\,{\rm s}$ for the cases of a drying droplet containing (d) only colloidal particles, (e) colloidal particles and motile bacteria, and (f) colloidal particles and non-motile bacteria. (d) Without bacteria and (f) with non-motile bacteria, the distribution decays fast away from zero and is characteristically asymmetric presenting a second smaller peak in the positive values indicating that the boundary height tend to increase, while, (e) with motile bacteria, the distribution is significantly broader and more symmetric around zero, indicating that the average height of the boundary can both increase or decrease (at least in the initial phases of the evaporation) as particles can be added to or removed from the boundary. See also supplementary video 2.
}
\label{fig3}
\end{figure*}

\begin{figure*}
\includegraphics[width=\textwidth]{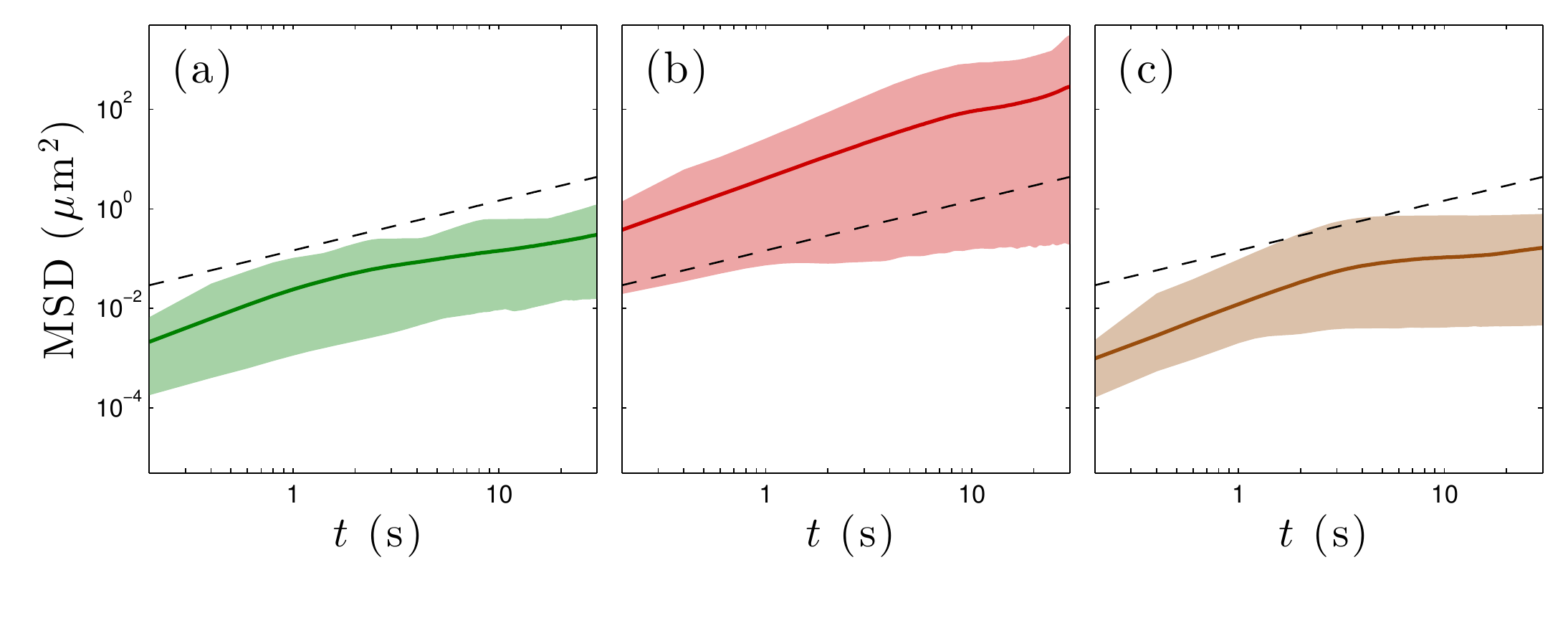}
\caption{
\textbf{Average mean square displacements (MSDs) of the particles after reaching the boundary.}
MSDs of $\sim 30$ particles from the moment they reach the boundary for the cases of a drying droplet containing (a) only colloidal particles, (b) colloidal particles and motile bacteria, and (c) colloidal particles and non-motile bacteria.
The solid lines represent the average MSDs and the shaded areas correspond to one standard deviation around the mean values. The black dashed lines represents the reference diffusive MSD for a Brownian particle in the bulk of an aqueous solution. (a) Without bacteria and (c) with non-motile bacteria, the motion of the particles is subdiffusive as the particles eventually settle within the boundary, as shown by the plateau at long time lags, while, (b) with motile bacteria, the particles perform active Brownian motion even after having reached the boundary, featuring  superdiffusive motion at small time differences and enhanced diffusion at longer time differences.
}
\label{fig4}
\end{figure*}

\begin{figure}[h!]
\includegraphics[width=0.45\textwidth]{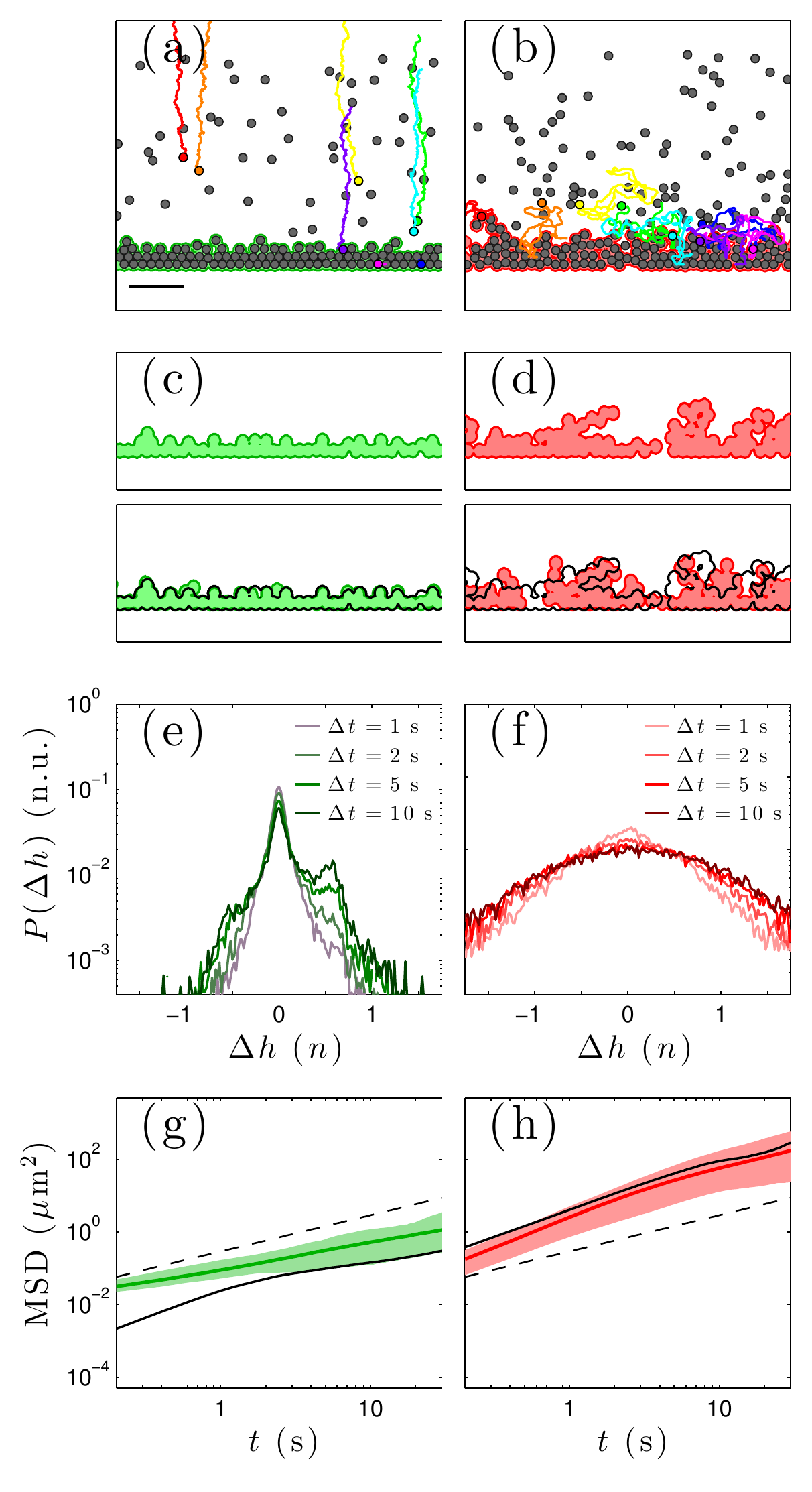}
\caption{
\textbf{Numerical simulations.}
The Brownian-dynamics simulation of the deposition of (a) passive and (b) active Brownian particles reproduces the features seen in Fig.~\ref{fig2}: (a) the particles get deposited along straight trajectories on the border which grows with crystalline close-packing for passive particles, while (b) active particles feature more complex trajectories, leading to a less ordered border. The scalebar corresponds to $20\,{\rm \mu m}$. (c-d) The boundary growth, (e-f) the histogram of the change of the boundary height, and (g-h) the mean square displacement (MSD) are also quantitatively similar to the experimental results provided in Figs.~\ref{fig3}(a-c), \ref{fig3}(d-f) and \ref{fig4}, respectively. In (g-h), the black solid lines represent the MSD measured experimentally in the cases without bacteria (Fig.~\ref{fig4}a) and with motile bacteria (Fig.~\ref{fig4}b), respectively. 
See also supplementary video 3.
}
\label{fig5}
\end{figure}

If a liquid droplet containing suspended particles is left to dry on a surface, it leaves behind a characteristic stain, which is often in the shape of a ring. This phenomenon is commonly observed when a drop of spilled coffee dries, hence the name {\em coffee ring}. In fact, this effect was already observed by Robert Brown in 1828 \cite{brown1829,goldstein2018}. The seminal work by Deegan and colleagues \cite{Deegan1997,Deegan2000,Deegan2000b} showed that coffee rings arise in a wide range of situations where the contact line of an evaporating droplet is pinned \cite{Hu2005,Marin2011,Yunker2013b,Weon2013,vanderKooij2016}. Following work has shown that the coffee ring effect can be controlled and tuned towards more uniform distributions via different approaches, e.g., by varying the size and shape of the suspended particles \cite{Weon2010,Yunker2011,Chasatia2011,Nguyen2013,Yunker2013} and by introducing static or dynamic Marangoni flows \cite{Hu2006,Li2013,Still2012,Sempels2013,Marin2016,Varanakkottu2016,Malinowski2018}. 

More recently, the coffee ring effect and its reversal have also been shown to impact microbial life \cite{Ott2004,Sempels2013,Yanni2017}. In fact, microorganisms can be exposed to environments that undergo occasional or periodic drying and, therefore, might make use of the coffee ring effect, or its absence, to optimise their chances of survival. For example, it has been shown that the production of biosurfactants by {\em Pseudomonas auriginosa} can reverse the coffee ring effect and produce more uniform stains upon drying \cite{Sempels2013}. Similarly, the coffee ring effect dictates interbacterial competition between different strains of {\em Vibrio cholerae} \cite{Yanni2017}.
Considering the fact that many bacteria are also motile to increase their survival chances \cite{Turnbull2001,Matz2005,hennes2017active}, a question naturally arises: What is the effect of self-motility on the dynamics occurring within a drying droplet?

In order to answer this question, we studied the drying process of droplets containing an active bath of motile bacteria (\emph{Escherichia coli} \cite{Berg2004}, wild-type strain RP437 from the \emph{E. coli} Stock Center at Yale University, cultured as described in \cite{Pince2016}) in the presence of suspended colloidal particles (polystyrene microspheres with diameter $2R = 3.00 \pm 0.07\,{\rm \mu m}$, initial volume fraction $\phi = 0.0055$). As a result of random collisions with the bacteria in the suspension, the colloids feature a crossover from superdiffusive motion at short times to enhanced diffusion at long times typical of self-propelled particles, such as bacteria and other microscopic active matter systems \cite{Morozov2014,Mino2011}. To keep the bacteria alive and motile during the evaporation process, droplets were made of motility buffer containing $10\,{\rm mM}$ monobasic potassium phosphate (${\rm KH_2PO_4}$), $0.1\,{\rm mM}$ EDTA, $10\,{\rm mM}$ dextrose, and 0.002\% Tween 20 \cite{Pince2016}.

At the beginning of each evaporation experiment we deposited a $0.6\pm0.04\,{\rm \mu l}$ droplet of this suspension on a glass microscope slide ($25.4\,{\rm mm} \times 76.2\,{\rm mm} \times 1\,{\rm mm}$), which had been previously cleaned with acetone and isopropanol, and washed with water after treatment with a 0.25 M NaOH solution for 5 minutes to make its surface more hydrophilic. The resulting droplet had a diameter of $2R_{\rm drop} = 2.5 \pm 0.2\,{\rm mm}$, a maximum height of $h_{\rm max} = 190 \pm 20\,{\rm \mu m}$, and a contact angle of $\theta_{\rm c} = 18 \pm 1^\circ$. 
The total evaporation time was $5 \pm 1$ minutes. 
We monitored the droplet evaporation using two imaging systems that projected the droplet on two CMOS cameras: the first imaging system with a $1.6\times$-magnification was used to record a video of the whole droplet's basal section at $5$ frames per second (fps), while the second with a $40\times$-magnification was used to record a magnified view of the dynamics at the droplet's edge at $5$ fps.

As a reference for the following results, Fig.~\ref{fig1}a shows the stain left behind by the evaporation of a droplet with only colloidal particles: because the motility buffer contains surfactants (i.e. Tween 20 and tryptone), the final stain show an irregular profile due to an increased probability of depinning events during the drying process \cite{Sefiane2004}. Under similar conditions of evaporation, Fig.~\ref{fig1}b shows that the addition of motile bacteria does not appreciably affect the uniformity of the resulting stain. The similarity between the two cases can be ascribed to the fact that the motility of the bacteria ($\sim 20\,{\rm \mu m\, s^{-1}}$) is significantly slower than the flows induced by the evaporation process (up to $\sim1\,{\rm mm\,s^{-1}}$), so that it does not play a role in influencing the deposition dynamics.

The speed of the bacteria and of the evaporation flows become comparable for slower evaporation rates ($25 \pm 5$ minutes). The evaporation rate was reduced by enclosing the evaporating droplet in a homemade chamber formed by two glass slides separated by a $400\,{\rm \mu m}$ parafilm spacer: This geometry allowed us to increase the local humidity in the droplet and thus extend its drying time in a controllable way up to 60 minutes. 
Under these slow evaporation conditions, the resulting stain is clearly different in the absence of the bacteria (Fig.~\ref{fig1}c) and in their presence (Fig.~\ref{fig1}d); in particular, the presence of motile bacteria produced a much more uniform {\em coffee disk} stain (see Supplementary Video 1) \cite{Yunker2013}. This is a first indication that, as the droplet evaporation rate slows down, the presence of active matter can alter the deposition of particles during the drying process and, thus, the resulting stain.

To get more insight on the influence of motility, we analysed the dynamics occurring at the edge of the droplet in the initial stages of the evaporation process \cite{Marin2011,Yunker2013b}, when the speeds of the flows and of the bacteria are most comparable.
Because of the presence of active matter, the edge growth process becomes more irregular, and the resulting stain shows less spatial order and lacks close packing. As shown in Fig.~\ref{fig2}a, the accumulation of particles at the border in the buffer solution without bacteria produces a border that is continuous and compact featuring hexagonal packing \cite{Xu2014}; furthermore, the trajectories in Fig.~\ref{fig2}a show that the particles go straight to the border due to the evaporation flows and, once reached it, they remain there forming a close-packed structure. Therefore, the boundary grows steadily with little rearrangement of the particles once they reach it (see left panel in supplementary video 2). 
As shown in Fig.~\ref{fig2}b, in the presence of motile bacteria, the resulting border is instead more jagged and loose with many empty regions as the particles form more complex structures with voids and no long-range order.
Perhaps more importantly, the boundary presents active dynamics: as can be seen from the trajectories of the particles in Fig.~\ref{fig2}b, after reaching the boundary, they still have a high probability of subsequently detaching from it pushed by the motile bacteria, and thus of rearranging the deposit under formation.

To check that these difference in dynamics was not due to the fact that the bacteria sterically impede the motion of the particles, we repeated the experiment with non-motile bacteria. In particular, we used bacteria that had stopped being motile as a consequence of the reduction of nutrients and oxygen in the buffer over time.
In this case, shown in Fig.~\ref{fig2}c (see also right panel in supplementary video 2), the microscopic profile of the border shows more irregularities than in the case without bacteria (Fig.~\ref{fig2}a), because of the presence of regions where non-motile bacteria accumulate. However, from the point of view of the edge growth dynamics, particles are deposited at the border and do not redetach, so that the area of the boundary increases steadily with no re-organisation of the position of the colloids within the edge. 

In order to make these qualitative observations more quantitative, we analysed the change in the edge height over time in the first 5 minutes from the initial formation of the boundary (Fig.~\ref{fig3}). At each instant, we determined the boundary shape by identifying the clusters of particles close to the edge of the droplet. Figs.~\ref{fig3}a-c show two such shapes at two instants separated by a time lag $\Delta t = 5\,{\rm s}$ (see also supplementary video 2) for the case without bacteria (Fig.~\ref{fig3}a), with motile bacteria (Fig.~\ref{fig3}b) and with non-motile bacteria (Fig.~\ref{fig3}c). While in Figs.~\ref{fig3}a and c, the boundary grows due to a steady deposition of additional particles over the previous boundary line, in Fig.~\ref{fig3}b the boundary is dynamically changing and rearranging over time as new particles arrive, thus reducing the similarity between the boundaries captured at different times. 
This information is quantitatively represented by the histograms of the spatial variation of the border height over time in Figs.~\ref{fig3}d-f. These histograms were calculated by measuring the change of the boundary area for different time lags $\Delta t = 1$, $2$, $5$ and $10\,{\rm s}$ to show how the border evolves over time. Without bacteria (Fig.~\ref{fig3}d) and with non-motile bacteria (Fig.~\ref{fig3}f), the distribution is narrow and skewed towards the positive side, confirming  that the border is steadily growing due to the addition of new particles that immediately reach their final position and then do not undergo further rearrangement. With motile bacteria (Fig.~\ref{fig3}e), the distribution is broader and more symmetric with similar likelihoods of both positive and negative values, indicating that the deposition of a particle and the escape of a particle are similarly likely within a short timeframe.

Qualitative information about the dynamics at the edge can also be extracted by analysing trajectories of individual particles once they reached the border for the cases without bacteria (Fig.~\ref{fig4}a), with motile bacteria (Fig.~\ref{fig4}b), and with non-motile bacteria (Fig.~\ref{fig4}c). An ideal tool to do this is to estimate their average mean square displacements (MSDs), which we calculated on 300s-long trajectories. For reference, the dashed lines in Figs.~\ref{fig4}a-c represent the theoretical MSD for a Brownian particle of the same dimension ($2R = 3\,{\rm \mu m}$) in the bulk of an aqueous solution. In the presence of motile bacteria (Fig.~\ref{fig4}b), the average MSD is about two orders of magnitude larger than in the absence of bacteria (Fig.~\ref{fig4}a) or in the presence of  non-motile bacteria (Fig.~\ref{fig4}c). In particular, the MSD in the presence of bacteria (red solid line in Fig.~\ref{fig4}b) shows superdiffusive behaviour (i.e. its slope is larger than 1) and is consistently above the reference of a passive Brownian particle (black dashed line); this is due to the fact that, after reaching the boundary, the particle on average leaves the boundary and behaves again as an active Brownian particle, before being dragged to the boundary again by the capillary flow, repeating this process a conspicuous number of times. This is strikingly different from the case without bacteria (Fig.~\ref{fig4}a) or with non-motile bacteria (Fig.~\ref{fig4}c), when the MSD shows subdiffusive behaviour and is below the reference MSD for a Brownian particle, indicating that, after an initial positional adjustment upon reaching the boundary, the particles are mainly stuck at a fix position.

In order to understand the dynamics of the boundary formation more deeply and to test its generality, we have developed a minimalistic numerical model capable of reproducing the key experimental observations and results. 
We describe the colloids in the drying droplet without bacteria as passive Brownian particles subject to a constant dragging force from the center to the edge of the droplet. We assume these particles to be hard-spheres to prevent particle overlapping.
As we are in an overdamped regime, we assume that the effect of this force on the dynamics of the particle is a constant drift velocity for each Brownian particle in the lower vertical direction.
We assume the boundary of the droplet to be a straight line in the horizontal direction and the dragging force to be perpendicular to the boundary in the vertical direction. 
We describe the particles in a droplet with bacteria as active Brownian particles, so that, in addition to the features described above, the particles also have a constant propulsion velocity in a direction that changes randomly on a time scale determined by an effective rotational diffusion time $\tau_{\rm r} = D_{\rm r}^{-1} \approx 5\,{\rm s}$, where $D_{\rm r}$ is the particle rotational diffusion \cite{Volpe2014}.
The results of the simulations have been analyzed following the same procedure described for the experimental acquisitions, and the outcome is summarised in Fig.~\ref{fig5} (see also Supplementary Video 3). The left column represents the case of passive Brownian particles, and the right column represents the case of active Brownian particles. Overall, the results of our simulations reproduce well the main experimental observations both in terms of histograms of the change of the boundary height and in terms of the average MSD of the particles after reaching the boundary.

In conclusion, we have demonstrated that the presence of active matter can dramatically influence the growth dynamics at the border of an evaporating droplet and the resulting stain. In particular, we have shown that the effects due to the presence of active matter arise when the drying dynamics occur on a time-scale comparable to the time-scales associated to the activity itself, and, therefore, they become particularly relevant for slowly drying droplets. Furthermore, our observation can help understand how bacteria and other motile microorganisms can influence their adaptation to environments that undergo periodic drying through their mobility. The possibility of influencing the coffee ring effect with active matter is not only relevant for several biophysical phenomena, such as the formation of biofilms and the spreading of pathogens in environments subject to periodic drying, but also for technological applications in printing \cite{Ikegawa2004}, coating \cite{Baldwin2011}, thin film deposition \cite{Zhang2015,Kaplan2015}, self-assembly of nanostructures \cite{Rabani2003,Suhendi2013,Ni2016}, where active matter could become an additional degree of freedom to dynamically control these processes \cite{Ni2013,Reichhardt2015}. 

The authors thank Naveed Mehmood for assistance with culturing the bacteria.
This work was partially supported by the ERC Starting Grant ComplexSwimmers (Grant No. 677511). 
We also acknowledge the COST Action MP1305 ``Flowing Matter" for providing several meeting occasions. PW was supported by an internship awarded by DAAD Rise.


\bibliographystyle{unsrt}

\begin{thebibliography}{}

\end{thebibliography}


\begin{thebibliography}{10}

\bibitem{brown1829}
R.~Brown.
\newblock additional remarks on active molecules.
\newblock {\em Phil. Mag.}, 6:161--166, 1829.

\bibitem{goldstein2018}
R.~E. Goldstein.
\newblock Coffee stains, cell receptors, and time crystals: Lessons from the
  old literature.
\newblock {\em Phys. Today}, 71:in press, 2018.

\bibitem{Deegan1997}
R.~D. Deegan, O.~Bakajin, T.~F. Dupont, G.~Huber, S.~R. Nagel, and T.~A.
  Witten.
\newblock Capillary flow as the cause of ring stains from dried liquid drops.
\newblock {\em Nature}, 389:827--829, 1997.

\bibitem{Deegan2000}
R.~D. Deegan, O.~Bakajin, T.~F. Dupont, G.~Huber, S.~R. Nagel, and T.~A.
  Witten.
\newblock Contact line deposits in an evaporating drop.
\newblock {\em Phys. Rev. E}, 62:756--765, 2000.

\bibitem{Deegan2000b}
R.~D. Deegan.
\newblock Pattern formation in drying drops.
\newblock {\em Phys. Rev. E}, 61:475--485, 2000.

\bibitem{Hu2005}
H.~Hu and R.~G. Larson.
\newblock Analysis of the microfluid flow in an evaporating sessile droplet.
\newblock {\em Langmuir}, 21:3963--3971, 2005.

\bibitem{Marin2011}
{\'A}.~G. Mar{\'\i}n, H.~Gelderblom, D.~Lohse, and J.~H. Snoeijer.
\newblock Order-to-disorder transition in ring-shaped colloidal stains.
\newblock {\em Phys. Rev. Lett.}, 107:085502, 2011.

\bibitem{Yunker2013b}
P.~J. Yunker, D.~J. Durian, and A.~G. Yodh.
\newblock Coffee rings and coffee disks: Physics on the edge.
\newblock {\em Physics Today}, 66(8):60--61, 2013.

\bibitem{Weon2013}
B.~M. Weon and J.~H. Je.
\newblock Self-pinning by colloids confined at a contact line.
\newblock {\em Phys. Rev. Lett.}, 110:028303, 2013.

\bibitem{vanderKooij2016}
H.~M. van~der Kooij, G.~T. van~de Kerkhof, and J.~Sprakel.
\newblock A mechanistic view of drying suspension droplets.
\newblock {\em Soft Matter}, 12:2858--2867, 2016.

\bibitem{Weon2010}
B.~M. Weon and J.~H. Je.
\newblock Capillary force repels coffee-ring effect.
\newblock {\em Phys. Rev. E}, 82:015305, 2010.

\bibitem{Yunker2011}
P.~J. Yunker, T.~Still, M.~A. Lohr, and A.~G. Yodh.
\newblock {Suppression of the coffee-ring effect by shape-dependent capillary
  interactions}.
\newblock {\em Nature}, 476:308--311, 2011.

\bibitem{Chasatia2011}
V.~H. Chasatia and Y.~Sun.
\newblock Interaction of bi-dispersed particles with contact line in an
  evaporating colloidal drop.
\newblock {\em Soft Matter}, 7:10135--10143, 2011.

\bibitem{Nguyen2013}
T.~A.~H. Nguyen, M.~A. Hampton, and A.~V. Nguyen.
\newblock Evaporation of nanoparticle droplets on smooth hydrophobic surfaces:
  The inner coffee ring deposits.
\newblock {\em J. Phys. Chem. C}, 117:4707--4716, 2013.

\bibitem{Yunker2013}
P.~J. Yunker, M.~A. Lohr, T.~Still, A.~Borodin, D.~J. Durian, and A.~G. Yodh.
\newblock Effects of particle shape on growth dynamics at edges of evaporating
  drops of colloidal suspensions.
\newblock {\em Phys. Rev. Lett.}, 110:035501, 2013.

\bibitem{Hu2006}
H.~Hu and R.~G. Larson.
\newblock Marangoni effect reverses coffee-ring depositions.
\newblock {\em J. Phys. Chem. B}, 110:7090--7094, 2006.

\bibitem{Li2013}
Y.-F. Li, Y.-J. Sheng, and H.-K. Tsao.
\newblock Evaporation stains: Suppressing the coffee-ring effect by contact
  angle hysteresis.
\newblock {\em Langmuir}, 29:7802--7811, 2013.

\bibitem{Still2012}
T.~Still, P.~J. Yunker, and A.~G. Yodh.
\newblock Surfactant-induced {M}arangoni eddies alter the coffee-rings of
  evaporating colloidal drops.
\newblock {\em Langmuir}, 28:4984--4988, 2012.

\bibitem{Sempels2013}
W.~Sempels, R.~De~Dier, H.~Mizuno, J.~Hofkens, and J.~Vermant.
\newblock Auto-production of biosurfactants reverses the coffee ring effect in
  a bacterial system.
\newblock {\em Nat. Commun.}, 4:1757, 2013.

\bibitem{Marin2016}
Alvaro Marin, Robert Liepelt, Massimiliano Rossi, and Christian~J K{\"a}hler.
\newblock Surfactant-driven flow transitions in evaporating droplets.
\newblock {\em Soft Matter}, 12:1593--1600, 2016.

\bibitem{Varanakkottu2016}
S.~N. Varanakkottu, M.~Anyfantakis, M.~Morel, S.~Rudiuk, and D.~Baigl.
\newblock Light-directed particle patterning by evaporative optical marangoni
  assembly.
\newblock {\em Nano Lett.}, 16:644--650, 2016.

\bibitem{Malinowski2018}
R.~Malinowski, G.~Volpe, I.~P. Parkin, and G.~Volpe.
\newblock Dynamic control of particle deposition in evaporating droplets by an
  external point source of vapor.
\newblock {\em J. Phys. Chem. Lett.}, 9(3):659--664, 2018.

\bibitem{Ott2004}
C.M. Ott, R.J. Bruce, and D.L. Pierson.
\newblock Microbial characterization of free floating condensate aboard the
  {M}ir space station.
\newblock {\em Microbial Ecol.}, 47:133--136, 2004.

\bibitem{Yanni2017}
D.~Yanni, A.~Kalziqi, J.~Thomas, S.~L. Ng, S.~Vivek, W.~C. Ratcliff, B.~K.
  Hammer, and P.~J. Yunker.
\newblock Life in the coffee-ring: How evaporation-driven density gradients
  dictate the outcome of inter-bacterial competition.
\newblock {\em arXiv}, 1707.03472, 2017.

\bibitem{Turnbull2001}
G.~A. Turnbull, J.~A.~W. Morgan, J.~M. Whipps, and J.~R. Saunders.
\newblock The role of bacterial motility in the survival and spread of
  \emph{{P}seudomonas fluorescens} in soil and in the attachment and
  colonisation of wheat roots.
\newblock {\em FEMS Microbiol. Ecol.}, 36:21--31, 2001.

\bibitem{Matz2005}
C.~Matz and K.~J\"{u}rgens.
\newblock High motility reduces grazing mortality of planktonic bacteria.
\newblock {\em Appl. Environmental Microbiol.}, 71:921--929, 2005.

\bibitem{hennes2017active}
M.~Hennes, J.~Tailleur, G.~Charron, and A.~Daerr.
\newblock Active depinning of bacterial droplets: The collective surfing of
  {\emph {b}acillus subtilis}.
\newblock {\em Proc. Natl. Acad. Sci. U.S.A.}, page 201703997, 2017.

\bibitem{Berg2004}
H.~C. Berg.
\newblock {\em \emph{E. coli} in Motion}.
\newblock Springer-Verlag, 2004.

\bibitem{Pince2016}
E.~Pin\c{c}e, S.~K.~P. Velu, A.~Callegari, P.~Elahi, S.~Gigan, G.~Volpe, and
  G.~Volpe.
\newblock Disorder-mediated crowd control in an active matter system.
\newblock {\em Nat. Commun.}, 7:10907, 2016.

\bibitem{Morozov2014}
A.~Morozov and D.~Marenduzzo.
\newblock Enhanced diffusion of tracer particles in dilute bacterial
  suspensions.
\newblock {\em Soft Matter}, 10:2748--2758, 2014.

\bibitem{Mino2011}
G.~Mi\~{n}o, T.~E. Mallouk, T.~Darnige, M.~Hoyos, J.~Dauchet, J.~Dunstan,
  R.~Soto, Y.~Wang, A.~Rousselet, and E.~Clement.
\newblock Enhanced diffusion due to active swimmers at a solid surface.
\newblock {\em Phys. Rev. Lett.}, 106:048102, 2011.

\bibitem{Sefiane2004}
K.~Sefiane.
\newblock Effect of nonionic surfactant on wetting behavior of an evaporating
  drop under a reduced pressure environment.
\newblock {\em J Colloid Interface Sci.}, 272:411--419, 2004.

\bibitem{Xu2014}
X.~Xu, L.~Ma, D.~Huang, J.~Luo, and D.~Guo.
\newblock Linear growth of colloidal rings at the edge of drying droplets.
\newblock {\em Colloids Surfaces A}, 477:28--31, 2014.

\bibitem{Volpe2014}
G.~Volpe, S.~Gigan, and G.~Volpe.
\newblock Simulation of the active {B}rownian motion of a microswimmer.
\newblock {\em Am. J. Phys}, 82:659--664, 2014.

\bibitem{Ikegawa2004}
M.~Ikegawa and H.~Azuma.
\newblock Droplet behaviors on substrates in thin-film formation using ink-jet
  printing.
\newblock {\em JSME Int. J. Ser. B Fluids Thermal Eng.}, 47:490--496, 2004.

\bibitem{Baldwin2011}
K.~A. Baldwin, M.~Granjard, D.~I. Willmer, K.~Sefiane, and D.~J. Fairhurst.
\newblock Drying and deposition of poly(ethylene oxide) droplets determined by
  {P}eclet number.
\newblock {\em Soft Matter}, 7:7819--7826, 2011.

\bibitem{Zhang2015}
X.~Zhang, A.~Crivoi, and F.~Duan.
\newblock Three-dimensional patterns from the thin-film drying of amino acid
  solutions.
\newblock {\em Sci. Rep.}, 5:10926--1--8, 2015.

\bibitem{Kaplan2015}
C.~N. Kaplan and L.~Mahadevan.
\newblock Evaporation-driven ring and film deposition from colloidal droplets.
\newblock {\em J. Fluid Mech.}, 781:1--13, 2015.

\bibitem{Rabani2003}
E.~Rabani, D.~R. Reichman, P.~L. Geissler, and L.~E. Brus.
\newblock Drying-mediated self-assembly of nanoparticles.
\newblock {\em Nature}, 426:272--274, 2003.

\bibitem{Suhendi2013}
A.~Suhendi, A.~B.~D. Nandiyanto, M.~M. Munir, T.~Ogi, L.~Gradon, and
  K.~Okuyama.
\newblock Self-assembly of colloidal nanoparticles inside charged droplets
  during spray-drying in the fabrication of nanostructured particles.
\newblock {\em Langmuir}, 29:13152--13161, 2013.

\bibitem{Ni2016}
S.~Ni, J.~Leemann, I.~Buttinoni, L.~Isa, and H.~Wolf.
\newblock Programmable colloidal molecules from sequential capillarity-assisted
  particle assembly.
\newblock {\em Science Adv.}, 2:e1501779, 2016.

\bibitem{Ni2013}
R.~Ni, M.~A. Cohen~Stuart, and M.~Dijkstra.
\newblock Pushing the glass transition towards random close packing using
  self-propelled hard spheres.
\newblock {\em Nat. Commun.}, 4:2704, 2013.

\bibitem{Reichhardt2015}
C.~Reichhardt and C.~J. Olson~Reichhardt.
\newblock Active microrheology in active matter systems: Mobility,
  intermittency, and avalanches.
\newblock {\em Phys. Rev. E}, 91:032313, 2015.

\end{thebibliography}

\end{document}